\documentclass{aa}
\usepackage{graphics,times}

\newcommand{\kms}{\hbox{km\,s$^{-1}$}}

\newcommand{\NH}{[N\,{\sc ii}]$\lambda$6583\AA/H$_{\alpha}$}

\begin{document}

\msnr{accepted version}

\title{An outflow from the nebula around the LBV candidate
S\,119}

\author{K.\ Weis \inst{1,2,}\thanks{Visiting Astronomer, Cerro
Tololo Inter-American Observatory, National Optical Astronomy
Observatories, operated by the Association of Universities for
Research in Astronomy, Inc., under contract with the National
Science Foundation.} \and W.J.\ Duschl \inst{1,2} \and D.J.\
Bomans \inst{3}}

\offprints{K.\ Weis, Heidelberg, Germany,
\email{kweis@ita.uni-heidelberg.de}}

\mail{K.\ Weis, Heidelberg, Germany}

\institute{Institut f\"ur Theoretische Astrophysik,
Tiergartenstr.\ 15, 69121 Heidelberg, Germany \and
Max-Planck-Institut f\"ur Radioastronomie, Auf dem H\"ugel 69,
53121 Bonn, Germany \and Astronomisches Institut,
Ruhr-Universit\"at Bochum, Universit\"atsstr. 150, 44780 Bochum,
Germany}

\date{Received / Accepted}

\abstract{We present an analysis of the kinematic and
morphological structure of the nebula around the LMC LBV
candidate S\,119. On HST images, we find a predominantly spherical
nebula which, however, seems to be much better confined in its
eastern hemisphere than in the western one. The filamentary
western part of the nebula is indicative of matter flowing out of
the nebula's main body. This outflow is even more evidenced
by our long-slit echelle spectra. They show that, while most of
the nebula has an expansion velocity of 25.5\,km\,s$^{-1}$, the
outflowing material reaches velocities of almost
140\,km\,s$^{-1}$, relative to the systemic one. A ROSAT HRI image
shows no trace of S\,119 and thus no indications of hot or shocked
material.
\keywords{Stars: evolution -- Stars: individual: S\,119 -- Stars:
mass-loss -- ISM: bubbles: jets and outflows}}
\maketitle

\section{Introduction}

Stars with masses in the range of $50 - 100\,{\rm M}_\odot$ and
luminosities of $10^5 - 10^6\,{\rm L}_\odot$ populate the upper
left end of the {\it Hertzsprung-Russell Diagram\/} (HRD). In
their short lives of $\la 10^7\,$yrs they evolve from hot O stars
on the main sequence towards cooler temperatures, first at almost
constant luminosities. They soon enter a phase of very strong
mass loss of up to $10^{-4}\,{\rm M}_\odot\,{\rm yr}^{-1}$. This
influences their further evolution dramatically: They do not
evolve further towards lower temperatures, i.e., towards the red
supergiant state, but rather turn in the HRD and become hotter
again, albeit later less luminous (e.g., Schaller et al.\ 1992;
Langer et al.\ 1994).

The region in the HRD where this turn occurs is known to be the
domain of the {\it Luminous Blue Variables\/} (LBVs). There exists
an empirical limit that separates a region in the HRD into which
the most massive stars do not evolve, the so-called {\it
Humphreys-Davidson Limit\/} (Humphreys \& Davidson 1979, 1994).
Here the stars not only exhibit large continuous mass loss, but at
least some of them undergo {\it giant eruptions\/}. Both, the
continuous wind and the eruptions lead to a peeling off of the
outer parts of the stellar envelope and to the formation of
circumstellar {\it LBV nebulae\/} (LBVN; e.g., Nota et al.\ 1995).
Humphreys \& Davidson (1994) classify 32 stars as LBVs and an
additional 9 as candidates. 9 of the LBVs and candidate stars are
located in the Milky Way and 10 in the {\it Large Magellanic
Cloud\/} (LMC).

S\,119 ($= {\rm Sk}$--69\,$175 = {\rm HDE}\,269687$) is one of the
LBVs in the LMC. It was classified as Ofpe/WN9 star by Bohannan
\& Walborn (1989). Since the early eighties, there was already
the suspicion of a close relation between Ofpe/WN9 stars and LBVs
when R127, located again in the LMC and previously classified as
Ofpe/WN9 underwent an LBV outburst (Stahl et al. 1983). The
evidence for a connection between the two stellar classes has
become even stronger since then, as longtime spectroscopic
monitoring of LBVs and Ofpe/WN9 stars became available (see,
e.g., Stahl \& Wolf 1986;  Wolf et al.\ 1988; Bohannan \& Walborn
1989; Nota et al.\ 1996; Pasquali et al. 1996).

After discovering a nebula around S\,119, Nota et al.\ (1994)
classified the star as an LBV candidate. Their ESO {\it New
Technology Telescope\/} (NTT) image shows a nebula of
7\arcsec\,$\times$\,9\arcsec\ size (corresponding to
1.9\,pc\,$\times$\,2.1\,pc for an assumed distance of the
LMC of 51.2\,kpc), with a
brighter lobe. Their NTT/EMMI spectra indicate an expansion
velocity of the S\,119 nebula of $\sim 25$\,\kms, and a ratio of
H$_\alpha$/N $\sim 1$, leading to \NH $\sim 0.75$. They derive a
radial velocity of the star and of the center of expansion in the
range of $100 - 140$\,\kms. This casts doubt on S\,119 being a
member of the LMC the radial velocity of which as derived from
H\,{\sc i} observations (Rohlfs et al.\ 1984) is typically in the
range of $240 - 300$\,\kms.

From the line ratio of [S\,{\sc ii}]6716/6731\,\AA\ Nota et al.\ (1994)
derived an electron density of $n_{\rm e} = 800$\,cm$^{-3}$
and---assuming an electron temperature of  $T_{\rm
e}=7\,500$\,K---estimated a nebula mass of $\sim 1.7$\,M$_\odot$.
Similar results for the nebula were reported by Smith et al.\
(1998). They describe the nebula as elliptical of size
7\farcs7\,$\times$\,8\farcs6, with $T_{\rm e} < 6\,800$\,K as estimated from
the non-detection of the [N\,{\sc ii}]5755\,\AA\ line, and
$n_{\rm e} = 680$\,cm$^{-3}$. Little reddening and a radial
velocity of $v_{\rm rad} = 118$\,\kms\ supports S\,119 not being a
member of the main body of the LMC. The main stellar parameters
of S\,119 have been determined by Crowther \& Smith (1997) using
two different models to account for the contamination of the
nebula in the stellar spectrum: $T_{\rm eff} = 26\,200\ /\
27\,000\,{\rm K}$, $L = 5.8\,10^5\ /\ 6.3\,10^5\,{\rm L}_\odot$,
and $\dot M = 1.34\,10^{-5}\ /\ 1.20\,10^{-5}\,{\rm
M}_\odot\,{\rm yr}^{-1}$.
In this contribution, we present results of an analysis of the
kinematics of the nebula around S\,119 and put it for the first
time into perspective with the nebula's high-resolution
morphology as obtained from {\it Hubble Space Telescope\/} (HST)
images. Moreover, we use the non-detection of S\,119 and its
nebula with the {\it High Resolution Imager\/} (HRI) on board the
{\it R\"ontgensatellit\/} (ROSAT) for determining and discussing
upper limits of the X-ray emission.

\section{Observation and data reduction}

\subsection{Imaging}

\begin{figure}
{\resizebox{\hsize}{!}{\includegraphics{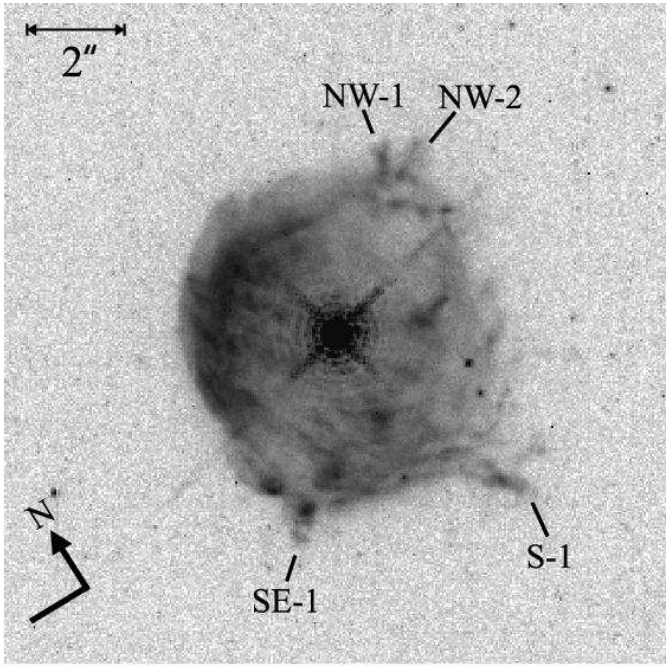}}} \caption{
The F656N HST image of S\,119 taken with the PC. The field of view
here is $\sim 15\arcsec\,\times\,15\arcsec$, and the intensity
scaling is logarithmic. A north-east vector indicates the
celestial orientation. The filaments described in Sect.\
\ref{sect:imaging} are designated as NW-1, NW-2, S-1, and SE-1.}
\label{fig:s119hst}
\end{figure}

\begin{figure}
{\resizebox{\hsize}{!}{\includegraphics{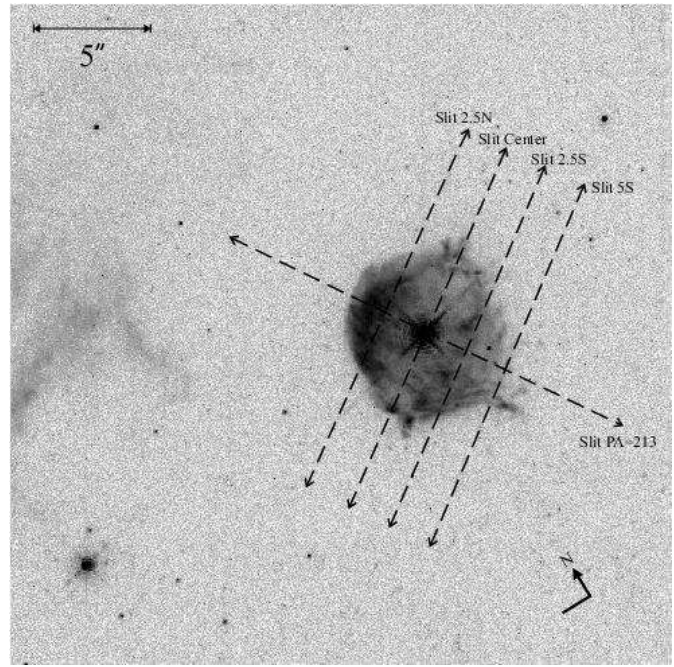}}} \caption{ A
larger area (30\arcsec\,$\times$\,30\arcsec) of the same HST image
as in Fig.\ \ref{fig:s119hst} shows the position and orientation
of the slits. In addition to the S\,119 nebula emission from a
closer H{\,\sc ii} region is visible.} \label{fig:hstslits}
\end{figure}

For the morphological analysis we retrieved from the STScI data
archive all images of S\,119 taken with the HST {\it Planetary
Camera\/} (PC) of the {\it Wide Field Planetary Camera 2\/} using
the F656N (H$_\alpha$) filter\footnote{program number: 6540; P.I.:
Regina Schulte-Ladbeck; dataset names: U4KY0301R...U4KY0308R}.
The exposure times were 500\,s for four images, and 5 and 30\,s
for two each. The data were reduced with the standard STSDAS/IRAF
routines. In total the four 500\,s exposures were combined and
cosmic-ray cleaned. They were not mosaiced since the nebula is
fully covered by the PC field of view. The pixel size in the
Planetary Camera is 0.0455\arcsec\ per pixel, the effective resolution about
0\farcs1.
The images were not rotated or binned to make sure to
maintain the full resolution. The
celestial directions therefore are indicated in the images. The
HST system position angle is 148.5\degr. Fig. \ref{fig:s119hst}
shows a section of $\sim 15\arcsec\,\times\,15\arcsec$ from the
reduced PC image which we used for the analysis. The almost full
field of view of the PC image is shown in Fig. \ref{fig:hstslits}.

\subsection{Long-slit echelle spectroscopy\label{sect:echelle}}

\begin{figure*}
\begin{center}
{\resizebox{15cm}{!}{\includegraphics{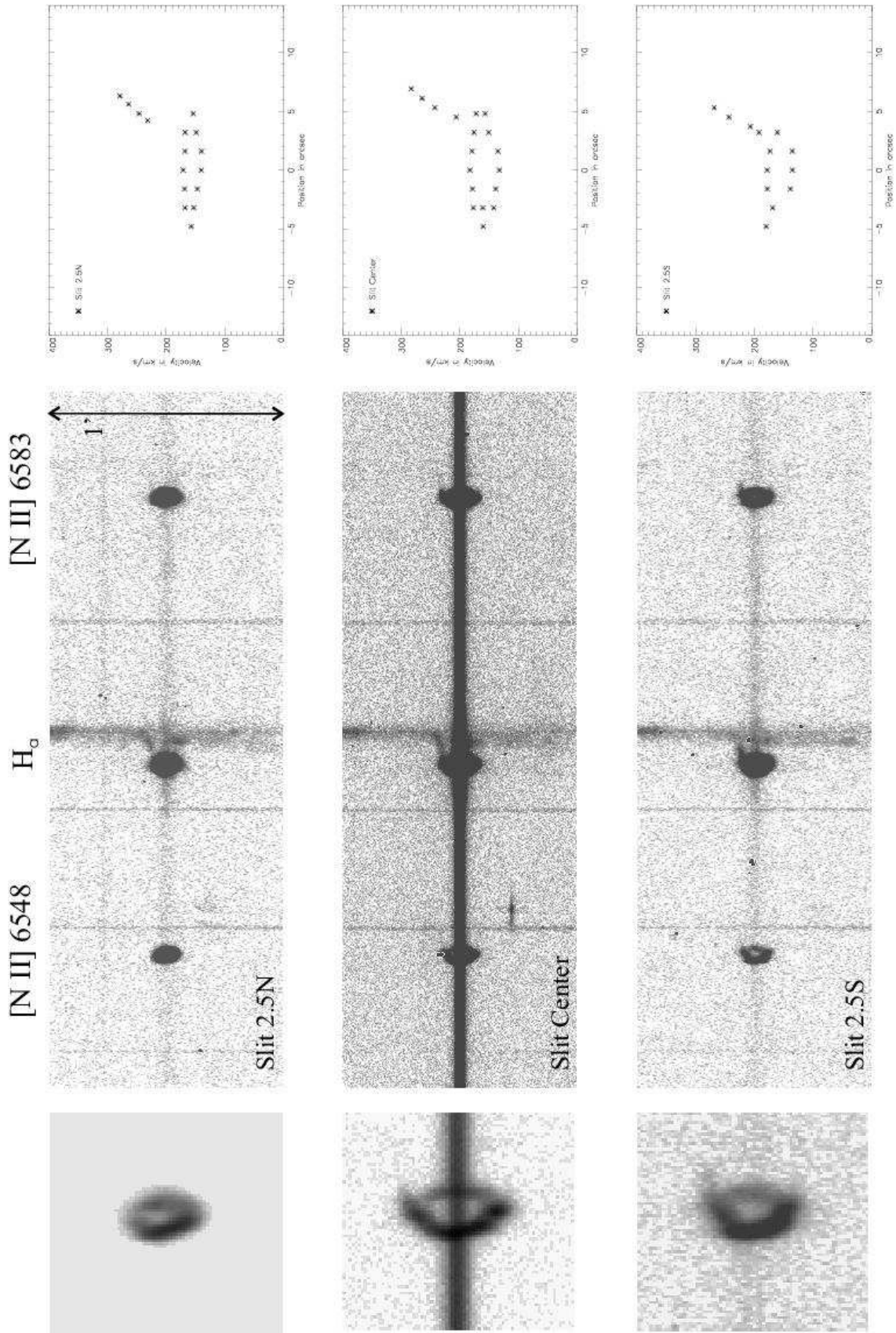}}}
\end{center}
\caption{Echellograms (center column) and corresponding
position-velocity diagrams (right column) for the observed slits.
Velocity measurements are with respect to the heliocentric system.
In left column an enlargment of [N\,{\sc ii}]$\lambda$6583\,\AA\
is shown.} \label{fig:echelle1}
\end{figure*}

\begin{figure*}
\begin{center}
{\resizebox{10.096cm}{!}{\includegraphics{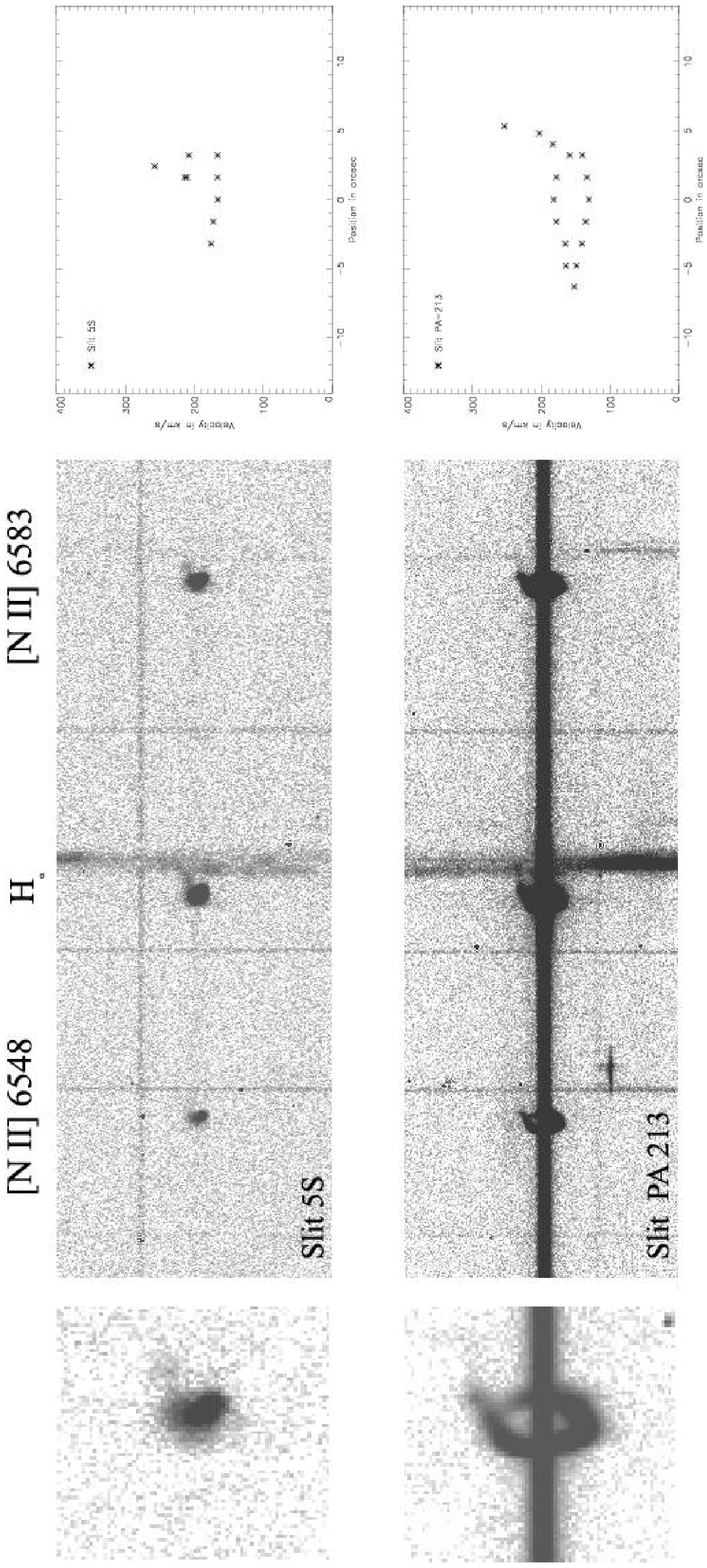}}}
\end{center}
\caption{Echellograms (center column) and corresponding
position-velocity diagrams (right column) for the observed slits.
Velocity measurements are with respect to the heliocentric system.
In left column an enlargment of [N\,{\sc ii}]$\lambda$6583\,\AA\
is shown.} \label{fig:echelle2}
\end{figure*}

For the kinematic analysis of the S\,119 nebula we obtained
high-resolution long-slit echelle spectra with the 4\,m-telescope
at the Cerro Tololo Inter-American Observatory. For the order
selection we replaced the cross-disperser by a flat mirror and
inserted a post-slit H$_\alpha$ filter (6563/75\,\AA). This setup
left us with a long-slit vignetted to a length of $\sim4^\prime$
and a spectral range that contained the H$_{\alpha}$ line and two
[N\,{\sc ii}] lines at 6548\,\AA\ and 6583\,\AA. We chose the
79\,l\,mm$^{-1}$ echelle grating and a slit-width of 150\,$\mu$m
(corresponding to 1\,\arcsec), which lead to an  instrumental FWHM
at the H$_\alpha$ line of 8\,km\,s$^{-1}$. The data were recorded
with the long focus red camera and a $2048\,\times\,2048$ pixel CCD
with a pixel size of 0.08\,\AA\,pixel$^{-1}$ along the dispersion
and 0$\farcs$26\,pixel$^{-1}$ on the spatial axis. Seeing was
$\sim 1\farcs2$. Thorium-Argon comparison lamp frames were taken
for wavelength calibration and geometric distortion correction.
Telluric lines visible in the spectra were used to improve the
absolute wavelength calibration of which we estimate an accuracy of
better than 0.08\,\AA.

We observed at two different position angles (PA) which were
nearly perpendicular to each other. Four slit positions mapped the
nebula with a ${\rm PA} = 125\degr$, one centered on the star {\it
(Slit Center)\/}, one offset 2\farcs5 to the north {\it (Slit
2.5N)} as well as one each at 2\farcs5 and 5\arcsec\ to the south
{\it (Slits 2.5S\/} and {\it 5S)\/}. One additional observation
was made with the slit oriented to ${\rm PA} = 213\degr$ and
centered on S\,119 {\it (Slit PA=213)\/}. Fig. \ref{fig:hstslits}
shows the slit positions. Fig. \ref{fig:echelle1} and
\ref{fig:echelle2} summarize our spectroscopic observations. The
left column of Fig. \ref{fig:echelle1} and \ref{fig:echelle2} show
the echellograms of each slit position we observed, the right
column the corresponding position velocity diagrams
($pv$-diagrams). The echellograms are 54\,\AA\ long and
extend 1\arcmin\ in spatial direction, centered on the projected
position of the star. The top of the echellograms points towards
north-west for ${\rm PA} = 125\degr$ and towards south-west for
${\rm PA} = 213\degr$. In each echellogram an insert (of
$20\arcsec\,\times\,5$\,\AA) shows the expansion
ellipse of the stronger [N\,{\sc ii}] line (6583\,\AA) at a
different cut level. The slits are identified according to the
nomenclature as introduced above. In the $pv$-diagrams the
positions are given as offsets from the projected location of the
star. Offsets to the north-west (or, in the case of Slit PA=213,
south-west) are counted positive. The velocity measurements for
the $pv$-diagrams followed from the H$_{\alpha}$ and the stronger
[N\,{\sc ii}] line at 6583\,\AA. We estimate an error of the
line fits to determine the radial velocities between $\pm$
0.5-1\,\kms, this is well below the printed symbol size in the
$pv$-diagramms. All velocities are with respect to the
heliocentric system, unless mentioned otherwise.

\subsection{ROSAT HRI data \label{sect:xray}}

In addition to the optical morphology and the kinematic analysis
we looked up ROSAT HRI observations of the S\,119 region. The
ROSAT satellite was sensitive to X-ray emission between 0.1 and
2.4\,keV and its HRI achieves a spatial resolution of $\sim
5\arcsec$. A 21\,ks ROSAT pointing\footnote{P.I.: You-Hua Chu,
HRI-pointing: rh600635n00} was retrieved from the {\it
Max-Planck-Institut f\"ur Extraterrestrische Physik\/} (MPE) ROSAT
data center. After reduction and analysis of the X-ray image with
IRAF/PROS\footnote{PROS is developed, distributed, and maintained
by the Smithonian Astrophysical Observatory, under partial support
from NASA contracts NAS5-30934 and NAS8-30751.}, we found no
traces of S\,119 in the data.

\section{Morphology from the HST image\label{sect:imaging}}

Previous images of the nebula around S\,119 taken with the NTT
show an axisymmetric shell, with a conspicuously brighter lobe at
the north east (Nota et al. 1994). The HST F656N images
resolve the nebula around S\,119 and show that it is approximately
spherically symmetric, with an average diameter of 7\farcs5.
Assuming that S\,119 is at the same distance as the LMC ($\sim
50$\,kpc), this corresponds to a size of 1.8\,pc. With the
help of the HST especially the small scale structures in and
around the nebula of S\,119 are clearly detected, which were not
seen in previous ground-based images. A large amount of these
filamemtary structures can be seen in addition to the spherical
main body of nebula. The most prominent ones are four filaments
extending out of the nebula, two next to each other to the
north-west (marked {\it NW-1\/} and {\it NW-2\/} in Fig.\
\ref{fig:s119hst}), one to the south ({\it S-1\/}), and one to the
south-east ({\it SE-1\/}). These filaments extend beyond the
nebula's main body by 0\farcs96, 1\farcs81, 0\farcs96, and
0\farcs68, respectively. On the west side, between filaments NW-2
and S-1, at the rim of the main body, numerous filaments of
comparatively low surface brightness extend beyond it. Here, no
clear border of the spherical main body is visible. In his region,
some of the filaments appear to be even detached from the main
body of the nebula.

In the east, the surface brightness of the nebula is highest, with
an H$_{\alpha}$ surface brightness of
8\,10$^{-14}$\,erg\,cm$^{-2}$\,s$^{-1}\,\sq^-\arcsec$. This
brighter area marks the region seen in seeing-limited NTT images.
In the inner part of the nebula the surface brightness is quite
low ($\sim 4\,10^{-14}$\,erg\,cm$^{-2}$\,s$^{-1}\,\sq^-\arcsec$
on average, with some of the filaments being weaker by almost
another order of magnitude). There it is far from a homogeneous
structure. On the other hand, there are also knots and filaments
distributed all over the nebula which are as bright as the
eastern part of the nebula. This combination of unevenly high and
low surface brightness features give rise to the nebula's patchy
appearance.

\section{The kinematic structure of the nebula}

Our high-resolution echelle spectra allow us to analyze the
kinematic structure of the S\,119 nebula in great detail. At a
FWHM resolution of 8\,\kms\ our observation fully resolve the
global structure of the Doppler ellipse of the expansion of the
S\,119 nebula.

\subsection{The overall expansion}

We find a radial velocity of the center of expansion at $156 \pm
2$\,\kms. This is in agreement with earlier findings that the
radial velocity of the S\,119 system is well below that of the
main part of the LMC. The echellograms show background
H$_{\alpha}$ emission along the entire slit at all positions,
which presumably results from an H\,{\sc ii} region within the
LMC. The radial velocity of this H\,{\sc ii} region of $\sim
264$\,\kms\ is consistent with it being located in the LMC. With
respect to this H\,{\sc ii} region, the S\,119 nebula moves with
about 100\,\kms\ more along the line of sight.

All spectra passing over the central part of the spherical main
body of the nebula show an expansion ellipse as expected for a
spherical expansion. We derive a maximum expansion velocity of
$25.5 \pm 2$\,\kms\ at the central position in both slits that
cross the central star (see $pv$-diagrams for Slit Center and for
Slit PA=213 at the position 0\arcsec\ in Figs. \ref{fig:echelle1},
\ref{fig:echelle2}).
As we move away from the geometric center of the nebula, the
slits (Slit 2.5N and Slit 2.5S) prove the decrease of the
expansion, that reduces to 15.5\,\kms\ and 21.4\,\kms\
respectively.

We measured sizes of the Doppler ellipses in our echelle data
and get diameters of 6\farcs0 (Slit 2.5N), 9\farcs1 (Slit Center),
and 6\farcs8 (Slit 2.5S), respectively. These sizes were
determined from the spectra and not from the binned $pv$-diagrams.
Given the considerably different spatial resolutions these
values agree with those derived from the HST images (5\farcs7,
8\farcs6, and 6\farcs2, respectively).

\begin{figure}
{\resizebox{\hsize}{!}{\includegraphics{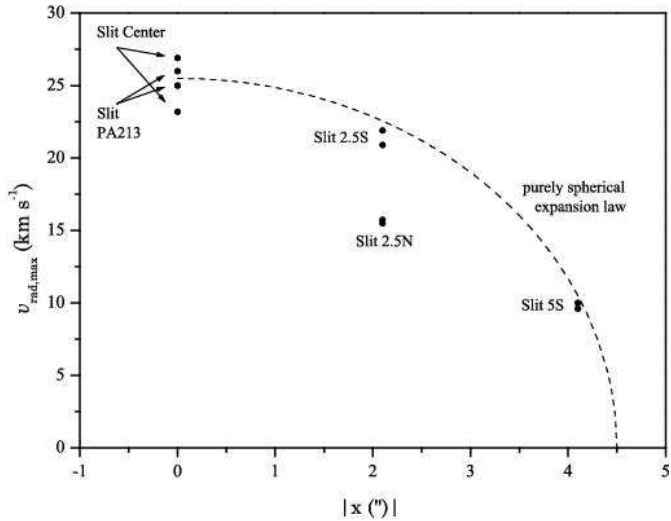}}} \caption{A
position-velocity diagram to compare the measurements with a model
of a spherically expanding shell (dashed curve). The errors
of the measurement are $\pm ( 0.5 - 1 )\,$ km\,s$^{-1}$, and thus
of about the size of the dot symbols or smaller \label{fig:comp}}
\end{figure}

For a spherical shell of radius $R$ a spherically symmetric
expansion with velocity $v_{\rm exp}$ would lead to an observed
radial velocity relative to the systemic velocity

\begin{equation}
v_{\rm rad,rel}^{\rm spher} (x,y) = v_{\rm exp} \left( \frac{R^2 -
x^2 - y^2}{R^2} \right)^{1/2}
\end{equation}
at a location $\left\{x,y\right\}$ in a Cartesian coordinate
system centered on the star. In the present case, we would assume
$R = 4\farcs5$ and $V_{\rm exp} = 25.5\,$\kms from our data. If we
orient the coordinate system such that the $y$-axis is along the
slit direction with $y = 0$ being the star's projected position
onto the slit and $x$ being the offset of the slit with respect to
the star, we get for the maximum radial velocities $v_{\rm rad,
rel, max} = v_{\rm rad, rel}^{\rm spher} (x=0,y) = 25.5\,\kms
\sqrt{1 - \left( y / 4\farcs5 \right)^2 }$. In Fig.\
\ref{fig:comp}\ we give a comparison between the absolute value of
the radial velocities (corrected for the systemic velocity) at the
star's position projected onto the slit, and the model values. If
it were a purely spherical expansion, the measured values should
fall onto the dashed model curve). We find a clear deviation from
a purely spherical expansion, for instance in the north-south
asymmetry in Slits 2.5N and 2.5S.

Both the red-shifted and the blue-shifted components of the
expansion ellipse have a FWHM of $0.3 - 0.5$\,\AA\ ($13 -
23$\,\kms), compared to an instrumental FWHM of 8\,km\,s$^{-1}$.
The echellograms in Fig. \ref{fig:echelle1} and \ref{fig:echelle2},
in particularly the
small inserts of the expansion ellipse reveal a clumpy
sub-structure. Most prominent in Slit Center, the blue-shifted
wing of the ellipse shows brightness variations (at a roughly
constant FWHM). These variations are most likely due to the
brighter knots identified in the HST image, which were
intercepted by the slits. However, due to the large
difference in resolution between our spectra and the HST image,
no clear identification of the individual knots is possible.

\subsection{The outflow}

\begin{figure}
{\resizebox{\hsize}{!}{\includegraphics{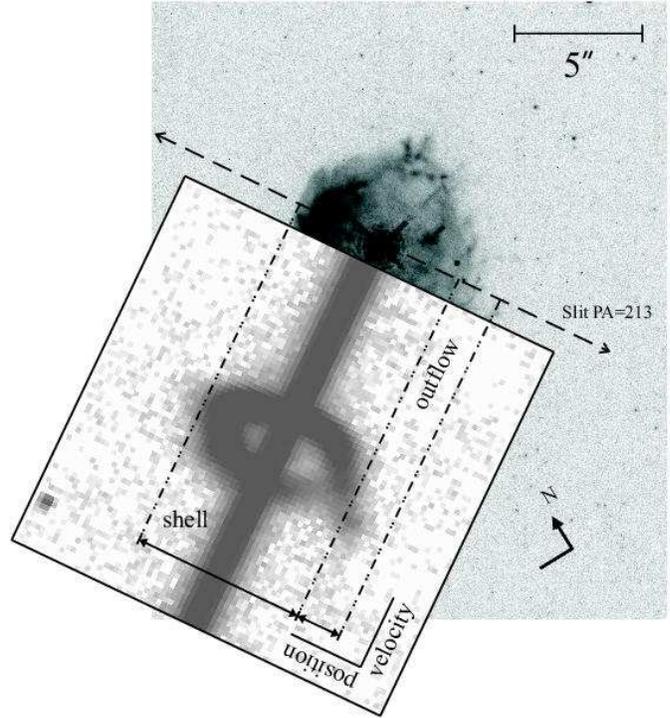}}}
\caption{Overlay of the HST image of S\,119 and the spectrum
Slit PA=213 around the [N\,{\sc II}]$\lambda$6583\AA line showing
the location of the outflow. \label{fig:outflow}}
\end{figure}

\begin{figure}
{\resizebox{\hsize}{!}{\includegraphics{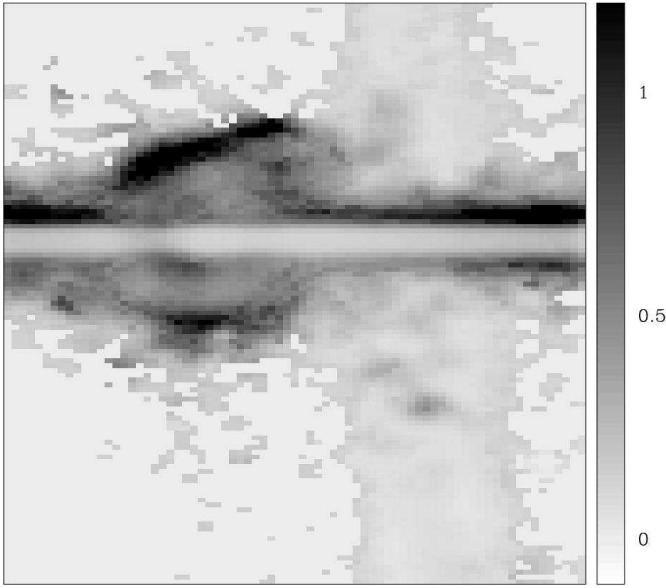}}}
\caption{Ratio map of the [N\,{\sc ii}]$\lambda$6583\AA\ and
H$_\alpha$ lines of spectrum Slit PA=213.  In this gray scale plot
of the two-dimensional spectrum we clipped the background so that
it is at a near noise-free zero value and slightly smoothed the
data with a $3\times 3$ median to improve the visibility of low
surface brightness features. Note the high \NH\ 
ratio of the ring and the outflow region compared
to the background H\,{\sc ii} region. \label{fig:ratio}}
\end{figure}

In addition to the spherical expansion we find in all spectra
clear indications for a much higher velocity expansion structure
(for an example, see Fig.\ \ref{fig:outflow}). It 
extends spatially from $\sim +4\arcsec$ to $\sim +7\arcsec$ and in
velocity space from the end of the expansion ellipse to more than
$280\,$\kms\ redshifted. Only in Slit 2.5N this emission feature
appears to be detached from the expansion ellipse, and merges with
the background H$_{\alpha}$ emission. To ensure that this high
velocity component is part of the S\,119 nebula and not a
structure in the background emission, we determined its [N\,{\sc
ii}]$\lambda$6583\AA/H$_\alpha$ ratio. If the feature contains CNO
processed material we expect a significantly larger value for an
LBV nebula than for an uncontaminated H\,{\sc ii} region. Nebulae
around LBVs do have a higher content of nitrogen due to CNO
processed material that is mixed into outer regions of the star,
which will be ejected and forms the nebula. The higher nitrogen
abundance of the nebula can be traced by the \NH\ ratio. Typical
examples of this ratio for LBVNs are $0.4 - 0.9$ for HR Car
(Hutsem\'ekers \& van Drom 1991; Weis et al.\ 1997), $0.7$ for AG
Car (Thackeray 1977; Smith et al.\ 1997), or even as high as $ 3 -
7$ in $\eta$ Car (Davidson et al.\ 1982; Meaburn et al.\ 1987;
1996; Weis et al.\ 1999). For the spherical nebula around S\,119
we derive a ratio of $0.6 \pm 0.1$, for the high velocity
component we still find $0.5 \pm 0.1$ (Fig.\
\ref{fig:ratio}). Both ratios are well within the range found in
LBVNs, and are significantly larger than the ratio of the H\,{\sc
ii} region in the background of which we observe [N\,{\sc
ii}]$\lambda$6583\,\AA/H$_\alpha = 0.04$. While not far from
the limit of our spatial resolution, the combination of spatial
and velocity resolution makes it unlikely that our \NH\
determination is contaminated. The \NH\  values derived for the
nebula agree well with  earlier measurements of Nota et al.
(1994).

The largest velocity of the high-velocity feature was detected in
the Slit Center position with 283.1\,\kms, almost 130\,\kms\
faster than the center of expansion of the central nebula. The
velocity of this component increases approximately linearly with
distance from the star's projected position. The echellograms do
not show any traces of this feature within the Doppler ellipse of
the spherical nebula. All these properties strongly point towards
this feature being a component of the nebula around S\,119 and
not a projection effect.

\begin{figure}
{\resizebox{\hsize}{!}{\includegraphics{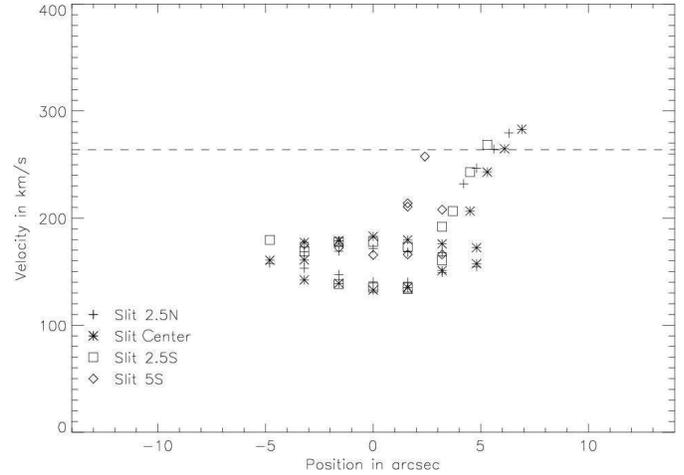}}} \caption{
Combination of all $pv$-diagrams with a PA=125\degr\ of Fig.\
\ref{fig:echelle1} and \ref{fig:echelle2}. The dashed line
indicates the radial velocity of the background LMC H\,{\sc ii}
region. } \label{fig:s119all}
\end{figure}

In Fig. \ref{fig:s119all} we combine the plots of all
$pv$-diagrams available to us. A dashed line marks the radial
velocity background H\,{\sc ii} region. Both, the spherical
expansion of the nebula's main body as well as the high-velocity
component are clearly visible. Due to its larger overall offset
from the star, in Slit 5S (diamond shape symbols) the outflow
seems to set on at a smaller projected offset from the star ($+
2\arcsec$ as compared to $+ 4\arcsec$ for the other slits). This
is caused by the effect that the slit intercepts the outflow
region earlier.

In Slit 2.5S (square shape symbols) at a projected position of $-
5\arcsec$ another feature appears which moves slightly faster by
$\sim 20\,{\rm km\,s}^{-1}$. This feature most likely can be
assigned to a part of the filament SE-1 identified in section
\ref{sect:imaging}.

\subsection{X-ray emission ?}

A 21\,ks integration with the ROSAT HRI shows no X-ray
emission of the nebula. At a distance of 50\,kpc and for a
Raymond-Smith plasma this corresponds to an upper flux limit of
$\sim 1\,10^{33}\,{\rm erg\,s}^{-1}$, almost independent of
assumed temperatures between 0.1 and 1\,keV (see Technical
Appendix to the ROSAT Call For Proposals, {\it ROSAT Appendix
F\/}). Thus we can only demonstrate that this limit is consistent
with the conditions in the S\,119 nebula:

\begin{itemize}
\item For an energy conserving bubble, for instance, we find an
expected X-ray flux of $\sim 2\,10^{32}\,{\rm erg\, s}^{-1}$
(Weaver et al.\ 1977, Chu et al.\ 1995; with a mechanical wind
luminosity of 8.1\,10$^{40}$\,erg\,s$^{-1}$, an electron density
of 700\,cm$^{-3}$, an electron temperature of 7\,000\,K, and an
expansion age of 2\,10$^4$\,yr).
\item Moreover, even for the highest outflow velocities ($\sim
10^2$\,km\,s$^{-1}$, may lead to post-shock temperatures at a
possible interface with the ambient medium of $T_{\rm ps} \approx
1.3\ 10^5\,{\rm K}$ and thus give rise to UV radiation rather than
X rays detectable by ROSAT.
\end{itemize}

\section{Summary and Conclusions}

We have analyzed the morphological and kinematic structure of the
nebula around the LBV candidate S\,119.

We find that S\,119 is surrounded by a nebula that represents a
spherically expanding shell. On the western rim of this shell, a
massive outflow occurs with a much higher velocity than the
expanding shell. If one assumes a scenario, in which LBV nebulae
are due to wind-wind interactions, the LBVN should be filled with
hotter gas, since LBVs---unless in outburst---are hot stars with
fast stellar winds. S\,119 is  a hot star, classified
as Ofpe/WN9. The lack of detectable X-rays can therefore be
explained by either {\it (a)\/} a relatively old outflow in which
the originally hot gas had already sufficient time to cool, or
{\it (b)\/} an LBV nebula which was not filled with hot gas in
the first place, or {\it (c)\/} a velocity of the outflow which
is too small to cause shocks with a sufficiently high post-shock
temperature, or---of course---a combination of two of these
reasons.

HST images are characterized by the spherical overall appearance
of this nebula. The edge of the spherical component is much
better defined in the northern and eastern directions where we
also find the largest surface brightness. It does not exhibit a
smooth distribution of surface brightness, but rather is very
patchy and shows many filaments and knots, some of which
extend---in particular in the western directions---beyond the
nebula's main body and occasionally are detached from the main
body. This is the reason for the less well defined edge of the
nebula in this direction.

High-resolution long-slit echelle spectra show that the spherical
component of the nebula is expanding with about 25\,\kms\ and that the
center of expansion is at a radial velocity of 156\,\kms. This is
remarkable as it is considerably lower than that of the LMC to
which---due to its projected position on the sky---the star seems
to belong. Both numbers are in good agreement with earlier
results (e.g., Nota et al.\ 1994).

In all spectra a high velocity component is present in addition to
the spherical expansion. It corresponds to a relative radial
motion of up to $\sim 100$\,\kms\ faster than the spherical main
body of the nebula. This high velocity material is concentrated on
the western side of the nebula. It is worth noting that on
this side the nebula also seems to be somewhat frayed even between
the filaments NW-2 and S-1 (see Sect.\ \ref{sect:imaging}). As
yet this makes S\,119 the only LBV (candidate) nebula with such an
outflow. The radial velocities in the outflow increase linearly
with distance from the nebula. This Hubble-type velocity law
reminds one of the strings found in and around the {\it
Homunculus\/} nebula of $\eta$ Car (Weis et al.\ 1999). However,
due to the rather different morphological structure, projection
effects (which were ruled out in the $\eta$ Car strings) may play
a role in the radial velocity structure in the S\,119 outflow. One
may, for instance, think of a conical structure with an opening
angle that increases outwards. Due to our current lack of
understanding of the physical nature of such Hubble type velocity
laws, it is not possible to draw firm conclusions on the relation
between the two phenomena. The largest radial velocity,
relative to the star, amounts to $127\,$\kms\ at a projected
distance of $6\farcs9$ from the star, corresponding to $\sim
1.7$\,pc. One may deduce a minimum dynamical age of $\sim
1.3\,10^4\,$yrs. The dynamical age of the shell assuming a radius
of 4\farcs5 and an expansion velocity of 25.5\,\kms amounts than
to $\sim 4.2\,10^4\,$yrs. One has to be aware, however, that it
depends strongly on the formation mechanism of the outflow how
meaningful a dynamical age is. In addition it was shown that the
radial velocity of the system seems to deviate from that of the
LMC so that the distance to S\,119 has to be questioned. This
distance on the other hand severely affects the determination of
the dynamical age. If S\,119 were only 30\,000\,pc away the
dynamical age of the nebula would already go down to $\sim
2.5\,10^4\,$yrs. That would than be comparable to the lifetime of
the star as an LBV. The dynamical age and the stars position in
the HRD, that is being a hot star makes it quite likely that
S\,119 is already on its way to leave the LBV phase as does He
3-519 (Davidson et al. 1993). 

The brightness difference between the west side and the east side
(Nota et al.\ 1994, see also Sect. \ref{sect:imaging}) can
most likely be accounted for by the outflow. However, one can only
speculate whether the outflow is due to a density gradient in the
ambient medium or whether it is caused by asymmetric flows in the
S\,119 system. Given the sphericity of the nebula's main component
we are inclined to hold environment effects responsible for the
outflow. This is also consistent with the brightest part of the
nebula occurring in the same direction in which the HST image
shows ionized diffuse gas which may easily indicate a higher
density in that direction.

\begin{acknowledgements}
We have made use of the ROSAT Data Archive of the
Max-Planck-Institut f\"ur extraterrestrische Physik (MPE) at
Garching, Germany. Part of the work was carried out on a
workstation provided by the {\it Alfried Krupp von Bohlen und
Halbach-Stiftung\/} to the ITA. This support is gratefully
acknowledged.
\end{acknowledgements}

\end{document}